\documentstyle[prb,aps]{revtex} %single space
\input epsf
\tighten
\begin{document}
\draft
\twocolumn[\hsize\textwidth\columnwidth\hsize\csname @twocolumnfalse\endcsname
\title{Static phase and dynamic scaling in a deposition model with an inactive
species}
\author{S. S. Botelho and F. D. A. Aar\~ao Reis}
\address{
Instituto de F\'\i sica, Universidade Federal Fluminense,\\
Avenida Litor\^anea s/n, 24210-340 Niter\'oi RJ, Brazil}
\date{\today}
\maketitle
\begin{abstract}
We extend a previously proposed deposition model with two kinds of particles,
considering the restricted solid-on-solid condition. The
probability of incidence of particle $C$ ($A$) is $p$ ($1-p$). Aggregation is
possible if the top of the column of incidence has a nearest neighbor $A$ and
if the difference in the heights of neighboring columns does not exceed $1$. 
For any value of $p>0$, the deposit attains some
static configuration, in which no deposition attempt is accepted. In $1+1$
dimensions, the interface width has a limiting value $W_s\sim p^{-\eta}$, with
$\eta = 3/2$, which is confirmed by numerical simulations. The dynamic scaling
relation $W_s=p^{-\eta}f\left( tp^z\right)$ is obtained in very large
substrates, with $z=\eta$.
\end{abstract}

\narrowtext
\vskip2pc]

\section{Introduction}

Statistical growth models of surfaces and interfaces have attracted
many attention in the last two decades motivated by technological applications
of thin films and related
nanostructures~\cite{barabasi,krug,meakin,halpinzhang}. In recent works,
models with two types of particles were introduced, in order to represent the
effects of different chemical species in the deposition
processes~\cite{wang,elnashar1,elnashar2,elnashar3,cn1,cn2,kotrla,drossel}.
The competition of different growth mechanisms may lead to crossover of growth
exponents and roughening transitions, as observed in many systems with a
single
species~\cite{alon1,alon2,gold,hinrichsen1,hinrichsen2,jricardo,albano1,albano2}.

A particularly interesting two-species model
was proposed by Wang and Cerdeira~\cite{wang}, which will be called $AC$
model. In that model, particles $A$ and $C$ are released with probabilities
$1-p$ and $p$, respectively, and aggregation is allowed only if the incident
particle encounters a neighboring $A$ at the sticking position (which
may be defined by different rules). Thus, particles $C$ represent impurities
that block the growth in their neighborhoods. For high $p$, the surface
will be contaminated with this species and the growth process will fail. In
previous works, the crossover of growth exponents was studied in the growth
regime~\cite{wang,elnashar1,elnashar2}.

In the present work, we will consider the restricted
solid-on-solid (RSOS) version of the $AC$ model. The RSOS model was introduced
by Kim and Kosterlitz in 1989 to describe the growth of thin films in which the
heights' differences between neighboring columns do not exceed certain
limiting value ${\Delta H}_{max}$.
This condition prevents the formation of high local
slopes in the film surface~\cite{kk}, then it is interesting for the
description of deposition processes in which diffusion and desorption
mechanisms (not explicitly included in the model) favor the
formation of locally smooth surfaces.

The RSOS version of the $AC$ model is defined as follows. At
each deposition attempt, an incident particle, $A$ or $C$, is chosen with the
probabilities $1-p$ and $p$, respectively. This particle is released above a
$d$-dimensional substrate in a randomly chosen column. The sticking position
for the incident particle is the top of the selected column, but the
aggregation is possible only if both the following conditions are satisfied:
(a) the difference in the heights of neighboring columns do not exceed ${\Delta
H}_{max}=1$; (b) the sticking position has a nearest neighbor particle $A$. If
one or both conditions are not satisfied, then the deposition attempt is
rejected. Fig. 1 illustrates the deposition rules.
\begin{figure}
\epsfxsize=8,6cm
\begin{center}
\leavevmode
\epsffile{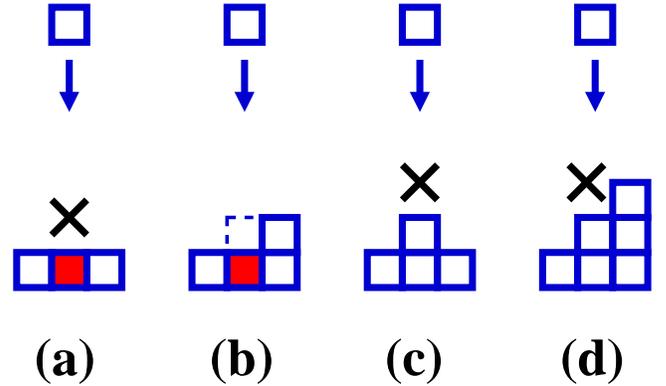}
\caption{Examples of application of the deposition rules of the RSOS version
of the $AC$ model. Open squares represent particles $A$ and filled squares
represent particles $C$. In (a), the aggregation attempt is not accepted
because there is no neighboring $A$ at the top of the column. In (b), this
neighbor is present (the dashed square indicates the sticking position). In (c)
and (d), the aggregation attempt is not accepted because it would violate the
RSOS condition.
}
\label{fig1}
\end{center}
\end{figure}

Here we will study the model in $d=1$.
We will show that a dynamic transition occurs at $p=0$
because any finite flux of particles $C$ will eventually suppress
the growth process. Thus, at $p>0$ the model presents a static phase, i. e.
the film attains a configuration that cannot continue growing because
no deposition attempt can be accepted. The interface width at saturation
scales with $p$ with an exponent $\eta$ that can be exactly obtained. It is
also shown that the dynamic exponent of the model is $z=\eta$.
The features of this static phase differ
from the dynamic nature of the smooth phases of other models with roughening
transitions, such as those including competition between adsorption and
desorption of adatoms~\cite{alon1,alon2,gold,hinrichsen1,hinrichsen2}.
However, there are many important open questions in the field of roughening
transitions, such as those concerning exponents' relations~\cite{alon2}, then
some results presented here may also be helpful in that context.

The rest of the paper is organized as follows. In Sec. II we present the
simulations' results and discuss the transition at $p=0$. In Sec. III, we
obtain a dynamic scaling relation for the interface width.
In Sec. IV we present our conclusions.

\section{Numerical simulations and the dynamical transition}

The main quantity of interest in deposition models is
the interface width $W$ of the deposit. In a surface of length $L$ ($L^d$
columns), at time $t$, $W$ is usually defined as
\begin{equation}
W\left( L,t\right) = {\left[ { \left< { { {1\over{L^d}} \sum_i{ {\left( h_i -
\overline{h}\right) }^2 } } } \right> } \right] }^{1/2} ,
\label{eq:1}
\end{equation}
where $h_i$ is the height of column $i$, the bar in $\overline{h}$ denotes a
spatial average and the brackets denote a configurational average, i. e., an
average over many realizations of the noise.

In the pure RSOS model ($p=0$), $W$ obeys the dynamic scaling relation
\begin{equation}
W \approx L^{\alpha} f\left( tL^{-z}\right) .
\label{eq:2}
\end{equation}
The exponents $\alpha$ and $z$ are consistent with
the Kardar-Parisi-Zhang (KPZ) theory~\cite{kpz}, which provides a hydrodynamic
description of kinetic surface roughening. In $d=1$, the KPZ equation gives
the exact values $\alpha = 1/2$ and $z=3/2$~\cite{kpz}. 

We simulated the RSOS version of the $AC$ model for several values of $p$,
most of them between $p=0.003$ and $p=0.02$. Substrates of lengths $L$ from
$L=256$ to $L=65536$ were considered, with periodic boundaries. During the
simulations, the time was measured as the number of deposition attempts per
column, thus one time unit corresponds to $L$ deposition attempts (accepted or
not). For each $p$ and $L$, we generated $10$ sets of ${10}^3$ different
deposits each one, and calculated error bars from the
fluctuations of the average values of the different sets.

In all cases, the growth process fails at sufficiently
long times, when the interface width attains a limiting value $W_s(p,L)$.
In Fig. 2 we show a deposit for $p=0.1$ and $L=128$ in which no
aggregation is possible. Notice that the deposit is faceted,
consisting of a set of droplets of triangular shape. In the valleys of the
deposit, there are triplets of particles $C$ with the structure shown in Fig.
3. Eventually, groups of four or more particles $C$ may create such valleys,
but they are much less probable then the triplets if $p$ is small. These
structures and the RSOS condition are responsible for the suppression of the
growth process.

For any $p>0$, $W_s$ converges to a finite value with vanishing $1/L$
corrections. It contrasts to the behavior of moving phases,
where the saturation width diverges as $L^{\alpha}$, with $\alpha >0$, in
agreement with Eq. (2). Despite this remarkable difference on finite-size
effects, $W_s$ will also be called saturation width here.
Extrapolations to $L\to\infty$ give
$W_s(p,\infty)$ and the average saturation height $H_s(p,\infty)$. The errors
in $H_s(p,\infty)$ are usually lower than $1\%$, and the errors in
$W_s(p,\infty)$ are nearly $10\%$.

\begin{figure}
\epsfxsize=8,5cm
\begin{center}
\leavevmode
\epsffile{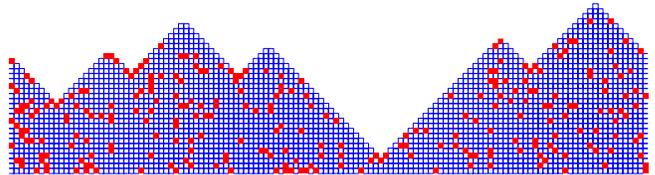}
\caption{Example of a final static deposit for $p=0.1$ and $L=128$.
}
\label{fig2}
\end{center}
\end{figure}

\begin{figure}
\epsfxsize=5,3cm
\begin{center}
\leavevmode
\epsffile{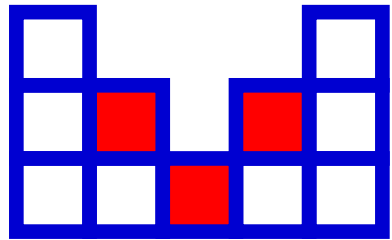}
\caption{Triplet of particles $C$ (filled squares), which occupies most
valleys of the static deposits, surrounded by $A$ particles.
}
\label{fig3}
\end{center}
\end{figure}

In Fig. 4a we plot $\log{H_s(p,\infty)}\times\log{p}$ and in
Fig. 4b we plot $\log{W_s(p,\infty)}\times\log{p}$. Those quantities scale as
\begin{equation}
W_s(p,\infty) \sim H_s(p,\infty) \sim p^{-\eta} ,
\label{eq:3}
\end{equation}
with $\eta = 1.509$ obtained from the least squares fit of the $H_s$ data, and
$\eta = 1.515$ obtained from the fit of the $W_s$ data. These relations show
that the growth process will actually fail for any $p>0$.

\begin{figure}
\epsfxsize=8,5cm
\begin{center}
\leavevmode
\epsffile{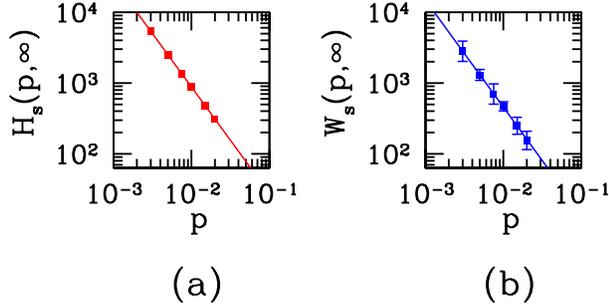}
\caption{(a) Saturation height $H_s(p,\infty)$ in very large substrates versus
probability $p$ of incidence of particles $C$; (b) Saturation width
$W_s(p,\infty)$ versus probability $p$. Solid lines are least squares fits.
}
\label{fig4}
\end{center}
\end{figure}

Our numerical results suggest the exact value $\eta = 3/2$, which can be
obtained using scaling arguments, as follows. The onset of triplets of
particles $C$ is responsible for the suppression of the growth process, and
each blocking configuration has probability of order $p^3$. A mound
of triangular shape (between valleys containing triplets of $C$) has height of
order $W_s$, then the number of particles $A$ in the mound is of order
${W_s}^2$. Thus, for small $p$, ${W_s}^2 \sim 1/p^3$, giving $\eta = 3/2$.

\section{Dynamic scaling}

The weak finite-size effects for large $L$ suggest that a dynamic scaling
relation in the static phase must be expressed only in terms of the probability
$p$ and the time $t$, while terms involving the length $L$ will be (vanishing)
corrections to scaling.

For very large $L$, we propose the scaling relation
\begin{equation}
W \approx p^{-\eta} f\left( t/\tau\right) \qquad ,\qquad \tau \sim p^{-z} , 
\label{eq:4}
\end{equation}
where $\tau$ is a characteristic time for the onset of correlations between
the $C$ triplets, and $z$ is a dynamic exponent.
$\tau$ is a measure of the number of layers of the deposit when these
correlations appear, thus we expect that $\tau\sim H_s$. Since $H_s$ also
scales with exponent $\eta$ (Eq. 3), we obtain
\begin{equation}
z = \eta = {3\over 2} ,
\label{eq:5}
\end{equation}

In order to test relation (5) with the above exponents $\eta$ and $z$, we
plot $W p^{\eta}$ versus $tp^z$ in Fig. 5, considering three values of
$p$: $p=0.005$, $p=0.01$ and $p=0.02$. Those data were obtained in substrates
with $L=65536$, which are sufficiently large to minimize finite-size effects.
The good data collapse in Fig. 5 confirms the validity of the scaling relation
(5).

\begin{figure}
\epsfxsize=8,5cm
\begin{center}
\leavevmode
\epsffile{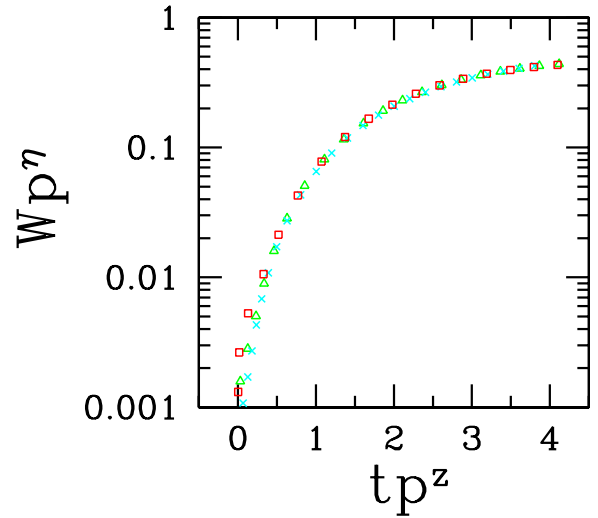}
\caption{Log-linear plot of $Wp^{\eta}$ versus $tp^z$, with $\eta =z=3/2$,
using data obtained in substrates with $L=65536$ and probabilities $p=0.02$
(squares), $p=0.01$ (triangles) and $p=0.005$ (crosses).
}
\label{fig5}
\end{center}
\end{figure}

Finally, it is interesting to notice the divergence of
the data for different $p$ at $t\lesssim 0.5p^{-z}$, as shown in Fig. 5. At
very short times, we expect that the interface width scales as in the pure
RSOS model, with no dependence on $p$, because the effects of $C$ particles
are negligible. Then, the pure RSOS model regime, in which the
width increases with time as $t^{1/3}$, becomes just a transient region for
any $p>0$.

\section{Discussion and conclusion}

We studied a deposition model with two types of particles, $A$ and $C$, in
which incident particles can only stick at positions that have a neighboring
$A$ and if the RSOS condition is satisfied. For any flux of particles $C$, the
growth eventually fails, due to the RSOS condition and the formation of
triplets of $C$. The saturation width $W_s$ is obtained
in the static final configurations in sufficiently large substrates.
Scaling arguments show that it scales as $W_s\sim p^{-3/2}$ for small $p$,
and this result is confirmed by numerical simulations. The interface width $W$
obeys a dynamic scaling relation involving the probability $p$ and the
deposition time $t$ (Eq. 4).

This model represents some growth mechanisms in the presence of
impurities. As proposed in Ref. \protect{\cite{wang}, it may describe the
effects of the deposition of an active particle $B$ that reacts with a
previously aggregated particle $A$ and forms the inactive particle $C$.
In the present RSOS version, small concentrations of the impurity may
suppress the growth process, with the inactive particles forming the pinning
centers. The blocking configurations depend on the particular model
considered (for instance, they will change for different ${\Delta H}_{MAX}$),
and the value of exponent $\eta$ depends on the number of particles $C$ in
those configurations. In a deposit with simple cubic lattice structure (which
is more suitable for real applications) and ${\Delta H}_{MAX}=1$,
configurations with five particles $C$ will form the pinning centers, and the
supression of the growth process will also be observed.

Previous works have also shown transitions from a moving phase to a smooth
phase~\cite{alon1,alon2,hinrichsen1,hinrichsen2}. Usually, the
roughening transitions are in the directed percolation (DP)
universality class, but some of them possess other symmetries, e. g. the
parity conserving class (PC). The smooth or anchored phases correspond to the
active (ordered) phases of DP, PC etc. In the moving phase, as the critical
points are approached, the growth velocities continuously decrease to zero.
The present model has many differences from those
ones. First, the growth velocity changes discontinuously from a finite value at
$p=0$ (pure RSOS model) to zero at $p>0$. Furthermore, if we consider the
order parameters $M_i$ defined in Ref. \protect{\cite{alon2} ($i=1,2,\dots$),
we obtain $M_i=0$ in the static phase, since there is no preferential level
for the pinning centers (see Fig. 2). Thus, this phase is not ordered
in that sense. Despite those
differences, we expect that the analysis that led to the dynamic scaling
relation (6) may be extended to other systems and may be useful to predict
relations between growth exponents.

\acknowledgements

This work was partially supported by CNPq and FAPERJ (Brazilian agencies).

\end{document}